\documentstyle[12pt]{article}
\title{\huge Order-disorder layering transitions of a spin-$1$ Ising model in a variable crystal field. } 
\author{\bf  
 L. Bahmad, A. Benyoussef, and H. Ez-Zahraouy$^*$ 
\\
 Laboratoire de Magn\'{e}tisme et de la Physique
 des Hautes Energies
\\
Universit\'{e} Mohammed V, Facult\'{e} des Sciences, Avenue Ibn Batouta,  B.P. 1014
\\Rabat, Morocco
}
\date{ }
\begin{document}
\maketitle 
\begin{abstract}
\mbox{  } The magnetic order-disorder layering transitions of a spin-$1$ Ising model are investigated, under the effect of a variable surface crystal field $\Delta_{s}$, using the mean field theory. Each layer $k$, of the film formed with $N$ layers, disorders at a finite surface crystal field distributed according to the law $\Delta_k=\Delta_s/k^\alpha$, $k=1,2,...,N$ and $\alpha$ being a positive constant.
We have established the temperature-crystal field phase diagrams and found a constant tricritical point and a reentrant phenomenon for the first $k_0$ layers. This reentrant phenomenon is absent for the remaining $N-k_0 $ layers, but the tricritical points subsist and depend not only on the film thickness but also on the exponent $\alpha$. On the other hand, the thermal behaviour of the surface magnetisation for a fixed value of the surface crystal field $\Delta_{s}$ and selected values of the parameter $\alpha$ are established.

\end{abstract}
%\vskip 1cm
\noindent
----------------------------------- 
\mbox{~}(*) Corresponding author: ezahamid@fsr.ac.ma \\
\mbox{~~} PACS :75.10.Jm; 75.30.Ds

\newpage

\section{Introduction}
\mbox{  } The Blume-Capel model (BC) was originally proposed to study the first-order magnetic phase transitions in spin$-1$ Ising systems [1].
This model was generalised to the Blume-Emery-Griffiths (BEG) to study phase separation and superfluity in $^3$He-$^4$He mixtures [2]. Later it has been applied to describe properties of multicomponent fluids [3], semiconductor alloys [4] and electronic conduction models [5]. The (BC) model is not exactly solvable in more than one dimension, but it has been studied over infinite $d$-dimensional lattices by means of many different approximate techniques and its phase diagram is well known. \\
When the theory of surface critical phenomena started developing, some attention has been devoted to the study of the (BC) model over semi-infinite lattices, with modified surface couplings. Benyoussef {et al.} [6] have determined the phase diagram in the mean field approximation, reporting four possible topologies at fixed bulk/surface coupling ratios. A similar analysis have also been done using a real space renormalization group transformation [7]. Other works referring to particular regions of the phase space are those using: the mean field approximation [8], the effective field approximation [9] and the low temperature expansion [10]. All these works show that it is possible to have a phase with ordered surface and disordered bulk, which is separated from the completely ordered phase by the so-called {\it extraordinary} transition and from the completely disordered phase by the {\it surface} transition. When such a phase is absent, the transition between the completely ordered and the completely disordered phase is called {\it ordinary}. The meeting point of the lines of these three kinds of phase transitions is named {\it special} and it is generally a multicrititcal point.
As discussed in Ref. [11], the strong interest in these models arises partly from the unusually rich phase transition behaviour they display as their interaction parameters are varied, and partly from their many possible applications. In most of these cases considered so far the bilinear interaction is ferromagnetic. In the antiferromagnetic case, the spin-$1$ Ising systems are used to describe both the order-disorder transition and the crystallisation of the binary allow, and it was solved in the mean field approach [12]. One of the most interesting and elusive features of the mean field phase diagram for the antiferromagnetic spin-$1$ Blume-Capel model in an external magnetic field is the decomposition of a line of tricritical points into a line of critical end points and one of double critical points [13]. This model was also studied by transfer-matrix and Monte Carlo finite-size-scaling methods [14], but such decomposition does not occur in this two dimensional model. 
On the other hand, ferroelectric films can be described by an Ising model and when the film becomes very thick, its properties are those of the semi-infinite Ising system [15-17]. From the experimental point of view, the most commonly studied magnetic multilayers are those of ferromagnetic transition metal such as Fe/Ni, where the coupling can exist between magnetic layers [18-20]. The discovery of enormous values of magnetoresistance in magnetic multilayers are far exceeding those found in single layer films and so exceeds the discovery of oscillatory interlayer coupling in transition metal multilayers. These experimental studies have motivated much theoretical works to study magnetic thin films as well as critical phenomena [21-26]. This is partly motivated by the development of new growth and characterisation techniques, but perhaps more so by the discovery of many exciting new properties, some quite unanticipated. 
Using the mean field theory, Benyoussef {\it et al.} [27] and Boccara {\it et al.} [28] have studied the spin-$1$ Ising model with a random crystal field. \\
The effect of the surface and bulk transverse fields on the phase diagrams of a semi infinite spin-$1$ ferromagnetic Ising model with a crystal field was investigated in [29] within a finite cluster approximation with an expansion technique for cluster identities of spin-$1$ localised spin systems. On the other hand, the transverse field or crystal field effects of spin-$1$ Ising model has been studied by several authors [30-33]. \\
The purpose of this paper is to study the effect of a variable crystal field according to the law $\Delta_k=\Delta_s/k^\alpha$ ($\Delta_s$ being the surface crystal field and $k$ the layer number from the surface and $\alpha$ a positive constant), on the order-disorder layering transitions of a spin-$1$ using the mean field theory. This paper is organised as follows. Section $2$ describes the model and the method. In section $3$ we present results and discussions. \\

\section{Model and method}
The experimental measurements of layer-by-layer ordering phenomena have been established on free-standing liquid crystals films such as ${\it nm}OBC$ (n-alkyl-4'-n-alkyloxybiphenyl-4-carboxylate) [34,35] and $54COOBC$ (n-pentyl-4'-n-pentanoyloxy-biphenyl-4-carboxylate) [36] for several molecular layers. More recently, Lin {\it el al.} [37] have used the three-level Potts model to show the existence of layer-by-layer ordering of ultra thin liquid crystal films of free-standing $54COOBC$ films, by adjusting the interlayer and intralayer couplings between nearest-neighbouring molecules.  \\
The system we are studying here is formed with $N$ coupled ferromagnetic square layers in the presence of a crystal field. The Hamiltonian governing this system is given by
\begin{equation}  
{\cal H}=-\sum_{<i,j>}J_{ij}S_{i}S_{j}+\sum_{i}\Delta_{i}(S_{i})^2
\end{equation}
where, $ S_{l}(l=i,j)=-1,0,+1$ are the spin variables. The interactions between different spins are assumed to be constant so that $J_{ij}=J$. The crystal field acting on a site $i$ is so that $\Delta_{i}=\Delta_{k}$ for all spins of the layer $k$ so that: $\Delta_{1}>\Delta_{2}>...>\Delta_{N}$. \\
Using the mean field theory, the quadrupolar moment and the magnetisation of a plane $k$ are given, respectively, by :
\begin{equation}    
q_{k}=<(S_{k})^2>=\frac{2\cosh(\beta h_{k})\exp(-\beta \Delta_{k})}{1+2\cosh(\beta h_{k})\exp(-\beta \Delta_{k})}
\end{equation} 
\begin{equation}    
 m_{k}=<S_{k}>=\frac{2\sinh(\beta h_{k})\exp(-\beta \Delta_{k})}{1+2\cosh(\beta h_{k})\exp(-\beta \Delta_{k})}
\end{equation}   
where $h_k=J(4m_{k}+m_{k+1}+m_{k-1})$, and $\beta=1/(k_B T)$, $k_B$ the Boltzmann constant ant $T$ the temperature. \\
The model we are studying, in this paper, corresponds to a crystal field distributed according to the law:
\begin{equation}    
\Delta_{k}=\Delta_{s}/k^{\alpha}
\end{equation}
where $\Delta_s=\Delta_{1}$ is to the crystal field acting on the surface (first layer $k=1$). Values of the parameter $\alpha$ will be discussed in the following; in particular $\alpha=0$ corresponds to a uniform crystal field $\Delta_s$ applied on each layer of the film. \\
The free energy of a layer $k$, can be expressed as:
\begin{equation}  
F_k=-\frac{1}{\beta}\log(1+2\cosh(\beta h_{k}))+\frac{1}{2}m_k(4m_{k}+m_{k+1}+m_{k-1})
\end{equation}
the parameters $h_{k}$ and $\beta$ still have the definitions given above. \\
As far as we know there is no natural compounds with decreasing crystal field from the surface to deeper layers. However, there exist a series of single crystals $R_{2}Fe_{14}B$ $(R=Y,Nd,Gd,Tb,Er,Tm)$ with a variable crystal field anisotropy. Indeed, by adjusting this anisotropy for each layer so that the amplitude of the crystal field will decrease from a layer $k$ to the next layer $k+1$. Hence, these materials can be grown layer-by-layer in a thin film with a decreasing crystal field from the surface to deeper layers. This can lead to an experimental realisation of our model.
Moreover, the decreasing function considered in this model, i.e. $k^{-\alpha}$, is not unique since the order-disorder layering transitions are found for any decreasing function of the crystal field from the surface to deeper layers.

\section{Results and discussion}
 \mbox{  }We are considering a system formed with $N$ ferromagnetic square layers of a simple cubic lattice spin$-1$ Ising model, with free boundary conditions. The notation $D^kO^{N-k}$ will be used to denote that the first $k$ layers are disordered while the remaining ${N-k}$ layers are ordered. In particular, $O^N$ corresponds to an ordered film whereas $D^N$ denotes a totally disordered film. The surface crystal field $\Delta_{k}$, applied on each layer $k$ is distributed according to the law given by Eq. $(4)$. \\
The ground state phase diagram of this system is illustrated by Fig. 1. 
For very small values of the surface crystal field $\Delta_s$, the system orders in the phase $O^N$. When increasing $\Delta_s$ the surface (first layer $k=1$) disorders and the phase $DO^{N-1}$ occurs at $\Delta_s/J=3(1)^\alpha$. Increasing $\Delta_s$ more and more, the second layer $k=2$ becomes disordered at $\Delta_s/J=3(2)^\alpha$, and so on. 
The transition from the phase $D^kO^{N-k}$ to the phase $D^{k+1}O^{N-(k+1)}$ is seen at $\Delta_s/J=3(k+1)^\alpha$ provided that $k+1 \le N$. 
For higher values of the surface crystal field the system is totally disordered and the phase $D^N$ occurs. \\
In particular, it is found that there exist a critical order layer $k_0$  corresponding to the transition $D^{k_0}O^{N-k_0} \leftrightarrow D^N$, at
a reduced surface crystal field given by:
\begin{equation}
\Delta_s/J=(3(N-k_0)-1)/(\sum_{k=k_0+1}^{N}(1/k^\alpha)).
\end{equation}
$k_0$ is exactly the number of layering transitions existing at $T=0$. 
It is shown that $k_0$ depends both on the parameter $\alpha$ and the film thickness $N$. The special case: $\alpha=0$ is a situation with a constant crystal field applied on each layer and there is only a single transition $O^N \leftrightarrow D^N$, occurring at $\Delta_s/J=3-1/N$ for $T=0$.\\
A preliminary study, for $N=2$, $3$, $5$, $10$ and $20$ layers, shows that the topology of the phase diagrams is not affected by the increasing film thickness, for a fixed exponent $\alpha$ value. Hence, the numerical results established in this work are done for a film with a thickness $N=10$ layers. \\
In order to outline the effect of the parameter $\alpha$ at a fixed film thickness, we plot in Figs. 2a, 2b, 2c and 2d the corresponding phase diagrams for several $\alpha$ values: $0.0$, $0.5$, $1.0$ and $2.0$. \\
In the case $\alpha=0.0$, Fig. 2a, a constant crystal field is applied on each layer and the corresponding transition $O^N \leftrightarrow D^N$ is found at $\Delta_s/J=3-1/N=2.9$ for very low temperatures.
Increasing $\alpha$ values, it is seen that large values of $\Delta_s/J$ are needed to disorder the film. This result is summarised in Figs. 2b, 2c and 2d. Indeed, due to the effect of the equation law $(4)$, separate layering transitions are found. For non null but very low temperatures, the surface crystal field values found in the phase diagrams are exactly those predicted by the ground state study: $\Delta_s/J=3(k)^\alpha$ for a layer order $k$ with $k \le k_0$. 
For example, in Fig. 2c for $\alpha=1.0$, the first transition is found at $\Delta_s/J=3.0$, the second transition at $\Delta_s/J=6.0$, the third transition at $\Delta_s/J=9.0$ and so on. 
The other numerical values of the surface crystal field corresponding to $\alpha=0.5$ in Fig. 2b, and $\alpha=2.0$ in Fig. 2d are exactly those predicted by the ground state phase diagram.
Furthermore, Figs. 2b, 2c and 2d show that the remaining $N-k_0$ transitions, not seen for $T=0$, arise for higher temperature values with separate tricritical points. More over, these results ensure the existence of a critical value $\alpha_c(k)$ above which a layer $k$ disorders at a very low temperature.  
The reentrant phenomena, seen in these figures for a given layer $k$, is caused by the competition of the temperature and an inductor magnetic field created by the deeper layers. Indeed, when the thermal fluctuations become sufficiently important, the magnetisation of some spins, of deeper layers, becomes non null (+1 or -1). This leads to the appearance of an inductor magnetic field. This magnetic field is responsible of the ordered phase seen for the layer $k$. This argument can also explain the absence of the reentrant phenomena for the last layer(s), once the magnetisation of the remaining layer(s) $N-k$ is not sufficient to create an inductor magnetic field. It is worth to note that the reentrant phenomena is always present for the layers $k$, ($k \leq k_0$), and the corresponding tricritical points $C_i$ are located at a constant temperature.
An interesting question arises at this stage: what is the effect of the parameter $\alpha$ on the surface of the film ($k=1$)? To clarify this point, we illustrate in Fig. 3a the profiles of the critical temperature for several values of $\alpha$ as a function of $\Delta_s/J$. 
It is found that increasing $\alpha$ values leads to important reentrants phenomena. Indeed, for large values of $\alpha$, the layer with $k > 1$ needs higher values of the surface crystal field to disorder. It is also found that, except the special case $\alpha=0.0$, the tricritical point is not affected by variations in $\alpha$ values. This surface tricritical point is located at $(\Delta_s/J=3.0,T_c/J=0.78)$ for $\alpha \neq 0$ and $(\Delta_s/J=2.9,T_c/J=1.9)$ for $\alpha=0$ for a film with a thickness $N=10$ layers. The effect of increasing $\alpha$ on a medium layer $k=N/2=5$ is illustrated in Fig. 3b. The same topology is found, but higher values of $\Delta_s/J$ are needed to overcome the disorder of this layer. \\
On the other hand, for $\alpha \rightarrow \infty$ the studied system is equivalent to a configuration with $\Delta_s/J(k=1)=cste $ and $\Delta_s/J(k > 1)=0.0$: the applied crystal field is present only at the surface and absent on the bulk. The corresponding phase diagram is illustrated by Fig. 4. 
Only two transitions are present in this case: $O^{10} \leftrightarrow DO^9$, occurring for low temperatures at $\Delta_s/J=3.0$, and $DO^9 \leftrightarrow D^{10}$ occurring at a constant temperature $T/J=3.899$ and any non null surface crystal field value.
In order to complete this study, we illustrate in Fig. 5, the thermal behaviour of the surface magnetisation for $\Delta_s/J=3.2$ and selected values of $\alpha$: 0.0, 0.2, 0.5, 1.0 and 2.0. At this surface crystal field value, the increasing $\alpha$ values effect is to increase both the surface magnetisation amplitude and the critical temperature needed to disorder this layer. This is a consequence of the factor $k^\alpha$ effect for $k > 1$, as it has been anticipated in the discussion above. \\

\section{Conclusion} We have studied the effect of a variable crystal field according to the law Eq. (4), and found order-disorder layering transitions of the Blume-Capel Ising model using mean field theory. We found a reentrant phenomena for each layer $k$ ($k \leq k_0$, see text), and a tricritical point at a fixed temperature with a vertical first order line.  We established the temperature-crystal field phase diagrams and found that the last layer tricritical point depend strongly on the small values of the exponent $\alpha$ for a fixed film thickness. 
The thermal behaviour of the surface magnetisation has also been investigated.

\newpage
\noindent{\bf References}
\begin{enumerate}

\item[[1]] M. Blume, Phys. Rev. {\bf 141}, 517 (1966); H. W. Capel, Physica {\bf 32}, 966 (1966). 

\item[[2]] M. Blume, V. J. Emery and R. B. Griffiths, Phys. Rev. A {\bf 4}, 1071 (1971). 

\item[[3]] J. Lajzerowicz and J. Sivardi{\'e}re, Phys. Rev. A {\bf 11}, 2079 (1975); J. Sivardi{\'e}re and J. Lajzerowicz, {\it ibid} {\bf 11}, 2090 (1975). 

\item[[4]] K. E. Newman and J. D. Dow, Phys. Rev. B {\bf 27}, 7495 (1983). 

\item[[5]] S. A. Kivelson, V. J. Emery and H. Q. Lin, Phys. Rev. B {\bf 42}, 6523 (1990). 

\item[[6]] A. Benyoussef, N. Boccara and M. Saber, J. Phys. C {\bf 19}, 1983 (1986).

\item[[7]] A. Benyoussef, N. Boccara and M. Bouziani, Phys. Rev. B {\bf 34}, 7775 (1986).

\item[[8]] X. P. Jiang and M.R. Giri, J. Phys. C {\bf 21}, 995 (1988).

\item[[9]] I. Tamura, J. Phys. Soc. Jpn {\bf 51}, 3607 (1982).

\item[[10]] C. Buzano and Pelizzola, Physica A {\bf 195}, 197 (1993).

\item[[11]] J.B. Collins, P.A. Rikvold and E. T. Gawlinski, Phys. Rev. B {\bf 38}, 6741 (1988).

\item[[12]] Y. Saito, J. Chem. Phys. {\bf 74}, 713 (1981).

\item[[13]] Y. L. Wang and K. Rauchwarger, Phys.Lett. A {\bf 59}, 73 (1976).

\item[[14]] J.D. Kimel, P. A. Rikvold and Y. L. Wang, Phys. Rev. B {\bf 45}, 7237 (1992).

\item[[15]] E. F. Sarmento, I. Tamura, L. E. M. C. de Oliveira and T. Kaneyoshi, J. Phys. C {\bf 17}, 3195 (1984).

\item[[16]] E. F. Sarmento and T. Kaneyoshi, Phys. Stat. Sol. b {\bf 160}, 337 (1990).

\item[[17]] Q. Jiang and Z. Y. Li, Phys. Rev. B {\bf 43}, 6198 (1991).

\item[[18]] N. M. Jennet and D. Dingley, J. Magn. Magn. Mater. {\bf 93}, 472 (1991).

\item[[19]] A. S. Edelstein {\it et al.}, Solid State Commun. {\bf 76}, 1379 (1990).

\item[[20]] R. Krishnan, H. O. Gupta, H. Lassri, C. Sella and J. Kaaboutchi, J. Appl. Phys. {\bf 70}, 6421 (1991).

\item[[21]] A. Z. Maksymowicz, Phys. Stat. Sol. b {\bf 122}, 519 (1984).

\item[[22]] A. Benyoussef and H. Ez-Zahraouy, Physica  {\bf 206 A}, 196 (1994).

\item[[23]] F. C. Sa Barreto and I. P. Fittipaldi, Physica A {\bf 129}, 360 (1985).

\item[[24]] T. Kaneyoshi Phys. Rev.  B {\bf 39}, 557 (1989).

\item[[25]] A. Z. Maksymowicz, Phys. Stat. Sol. b {\bf 122}, 519 (1984).

\item[[26]] A. Bobak, Phys. Stat. Sol. b {\bf 109}, 161 (1982).

\item[[27]] A. Benyoussef, T. Biaz, M. Saber and M. Touzani, J. Phys. C {\bf 20}, 5349 (1987).

\item[[28]] A. Benyoussef and H. Ez-Zahraouy, J. Phys.: Cond. Matt.  {\bf 6}, 3411 (1994).

\item[[29]] N. Cherkaoui Eddeqaqi, H. Ez-Zahraouy and A. Bouzid, J. Magn. Magn. Mat. {\bf 207}, 209 (1999).

\item[[30]] R. B. Stinchcombe, J. Phys. C {\bf 6}, 2459 (1973).

\item[[31]] J. L. Zhong, J. Liangli and C. Z. Yang, Phys. Stat. Sol. (b) {\bf 160}, 329 (1990).

\item[[32]] U. V. Ulyanoc, O. B. Zalavskii, Phys. Rep. {\bf 216}, 179 (1992)

\item[[33]] A. Benyoussef, H. Ez-Zahraouy and M. Saber, Phys. A {\bf 198}, 593 (1993)

\item[[34]] R. Geer, T. Stoebe, C. C. Huang, R. Pindak, J. W. Goodby, M. Cheng, J. T. Ho and S. W. Hui, {\it Nature} {\bf 355}, 152 (1992).

\item[[35]] T. Stoebe, R. Geer, C. C. Huang and J. W. Goodby, Phys. Rev. Lett. {\bf 69}, 2090 (1992).

\item[[36]] A. J. Jin, M. Veum, T. Stoebe, C. F. Chou, J. T. Ho, S. W. Hui, V. Surendranath and C. C. Huang, Phys. Rev. Lett. {\bf 74}, 4863 (1995); Phys. Rev. E {\bf 53}, 3639 (1996).

\item[[37]] D. L. Lin, J. T. Ou, Long-Pei Shi, X. R. Wang and A. J. Jin, Europhys. Lett. {\bf 50}, 615 (2000).

\end{enumerate}

\newpage
\noindent{\bf Figure Captions}\\

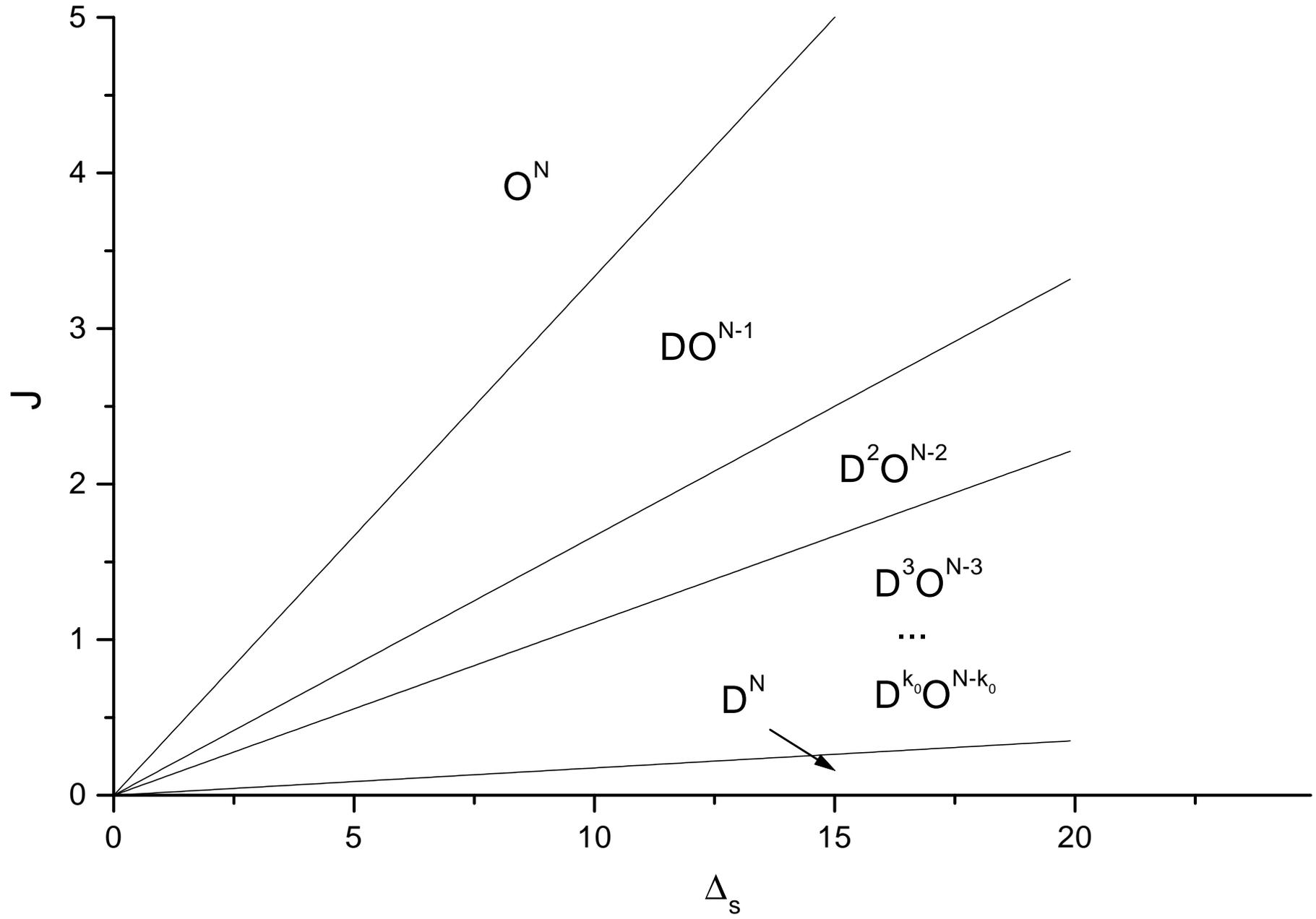
\noindent{\bf Figure 1.}: The ground state phase diagram in the plane $(J,\Delta_s)$ for a film thickness $N$. The notations $D^{k}O^{N-k}$ are defined in the body text. The calculations presented in this manuscript correspond to a film thickness $N=10$ layers. \\

\noindent{\bf Figure 2.}: The critical temperature behaviour as a function of the surface crystal field $\Delta_s$ for several values of the exponent $\alpha$ : 
a) $\alpha=0.0$,  
      b) $\alpha=0.5$, 
      c) $\alpha=1.0$ and
 d) $\alpha=2.0$. \\
The first-order transition lines  (dashed line) are connected to the second-order transition lines (continuous line) by a tricritical point. \\

\noindent{\bf Figure 3.}: The critical temperature behaviour of the first layer $k=1$ and the medium layer $k=5$, as a function of the surface crystal field $\Delta_s$ for selected values of the parameter $\alpha$: $0.0, 0.2, 0.5 1.0$ and $2.0$. 
      a)  First layer $k=1$.  
      b)  Medium layer $k=N/2=5$. \\

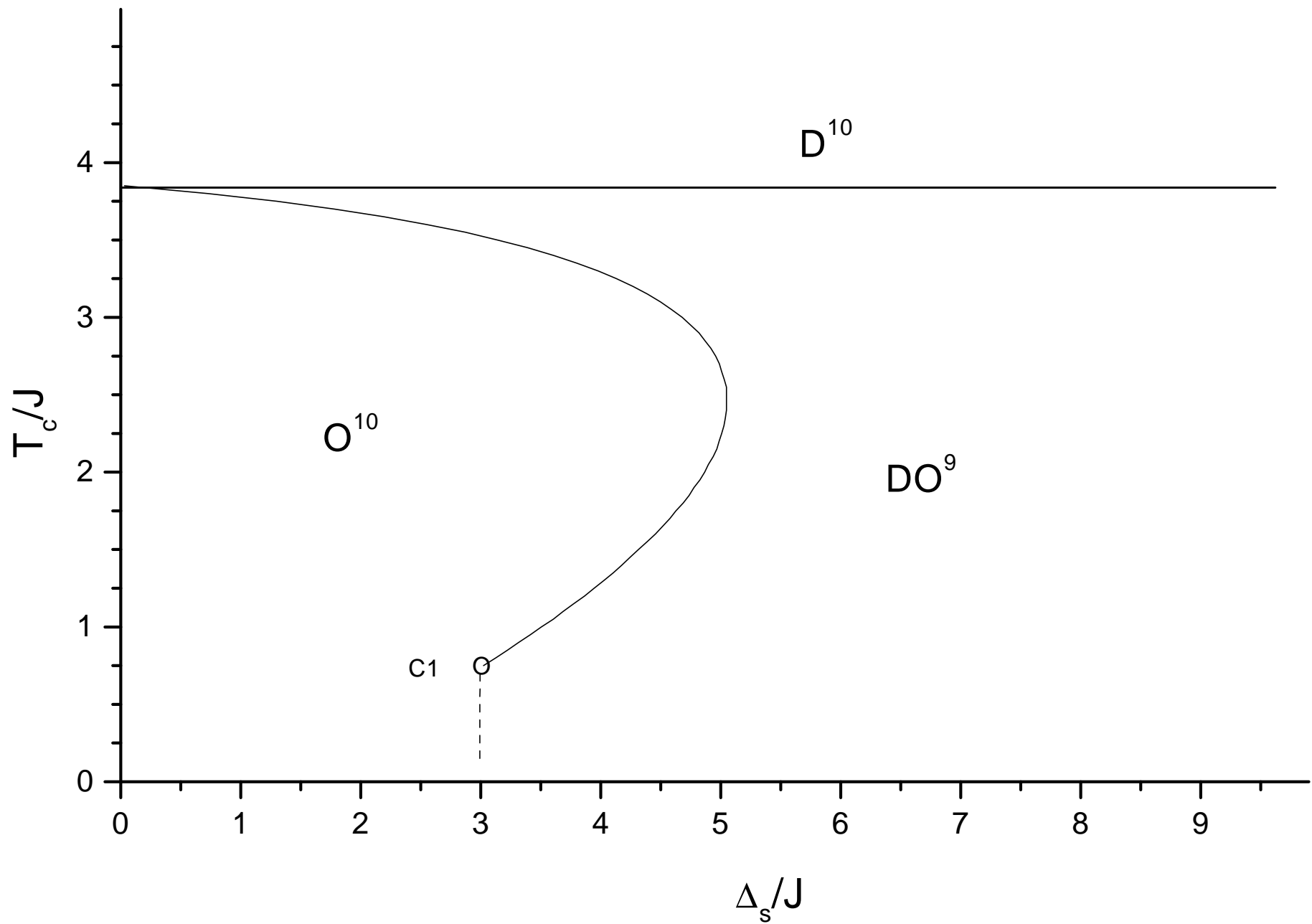
\noindent{\bf Figure 4.}: The dependence of the critical temperature as a function of the surface crystal field $\Delta_s$ for $\alpha=\infty$. Only two transitions are present in this case: $O^{10} \leftrightarrow DO^9$ and $DO^9 \leftrightarrow D^{10}$. \\  

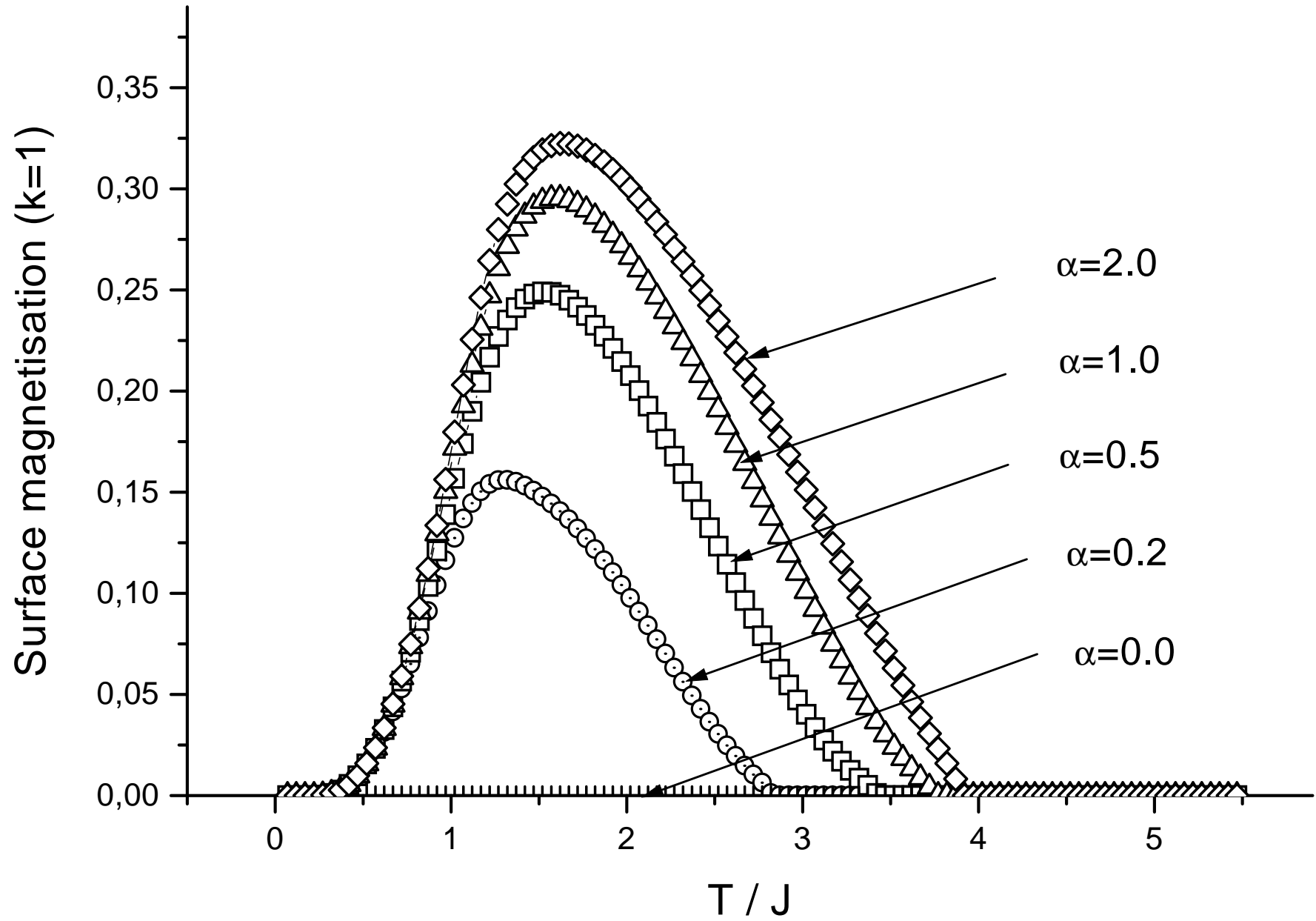
\noindent{\bf Figure 5.}: The thermal behaviour of the surface (layer $k=1$) magnetisation for selected values of $\alpha$: $0.0, 0.2, 0.5, 1.0$ and $2.0$.   \\

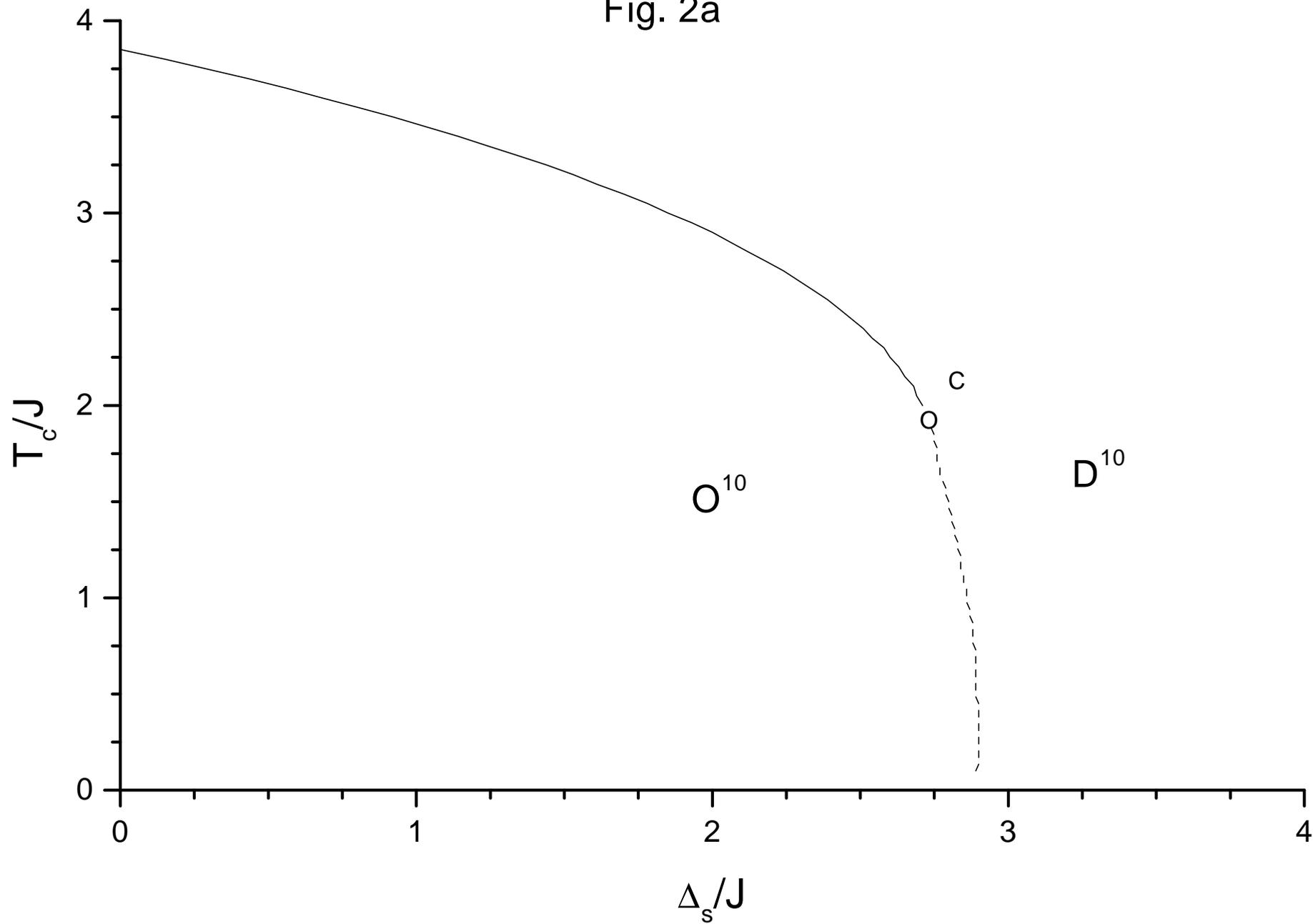
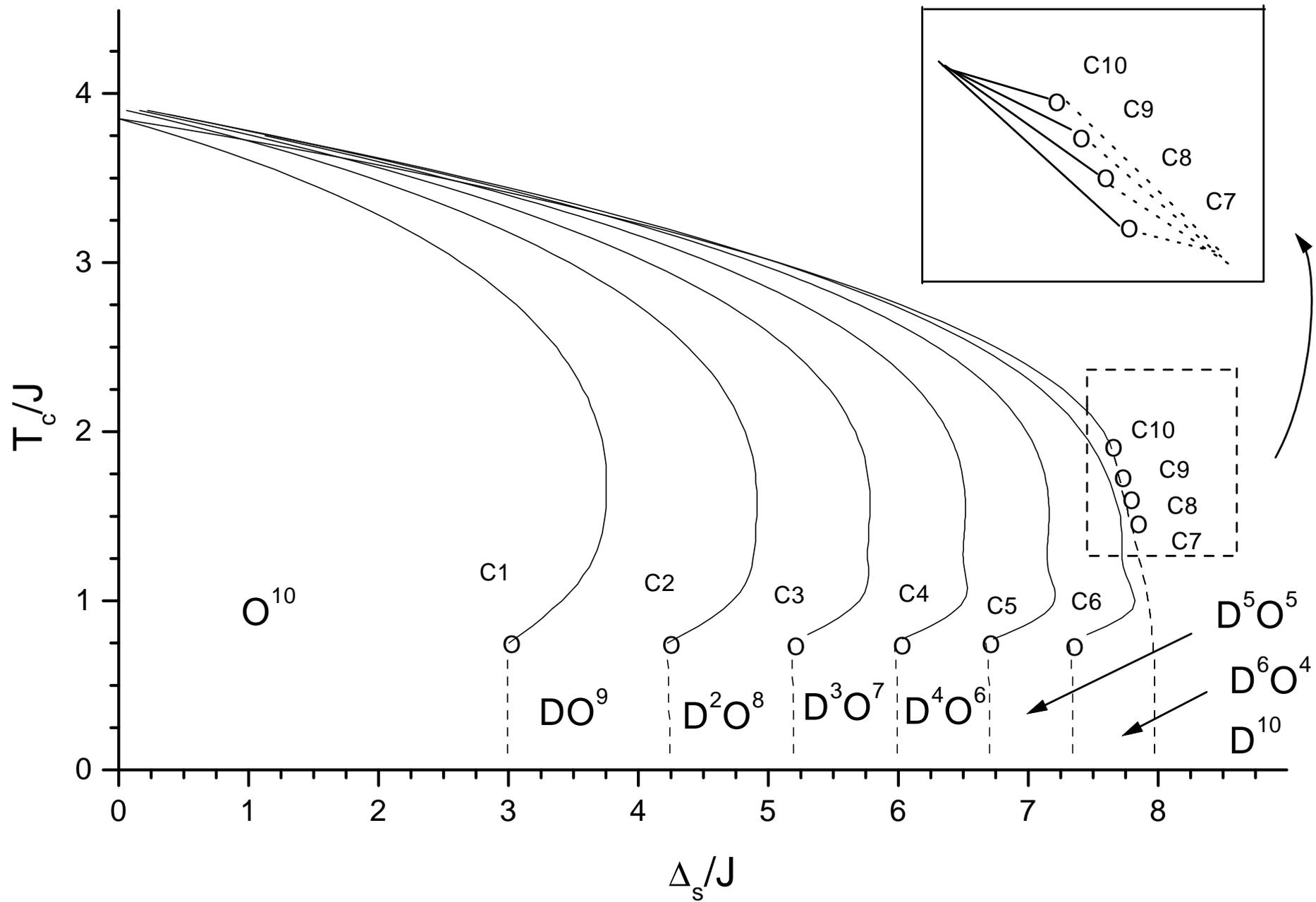
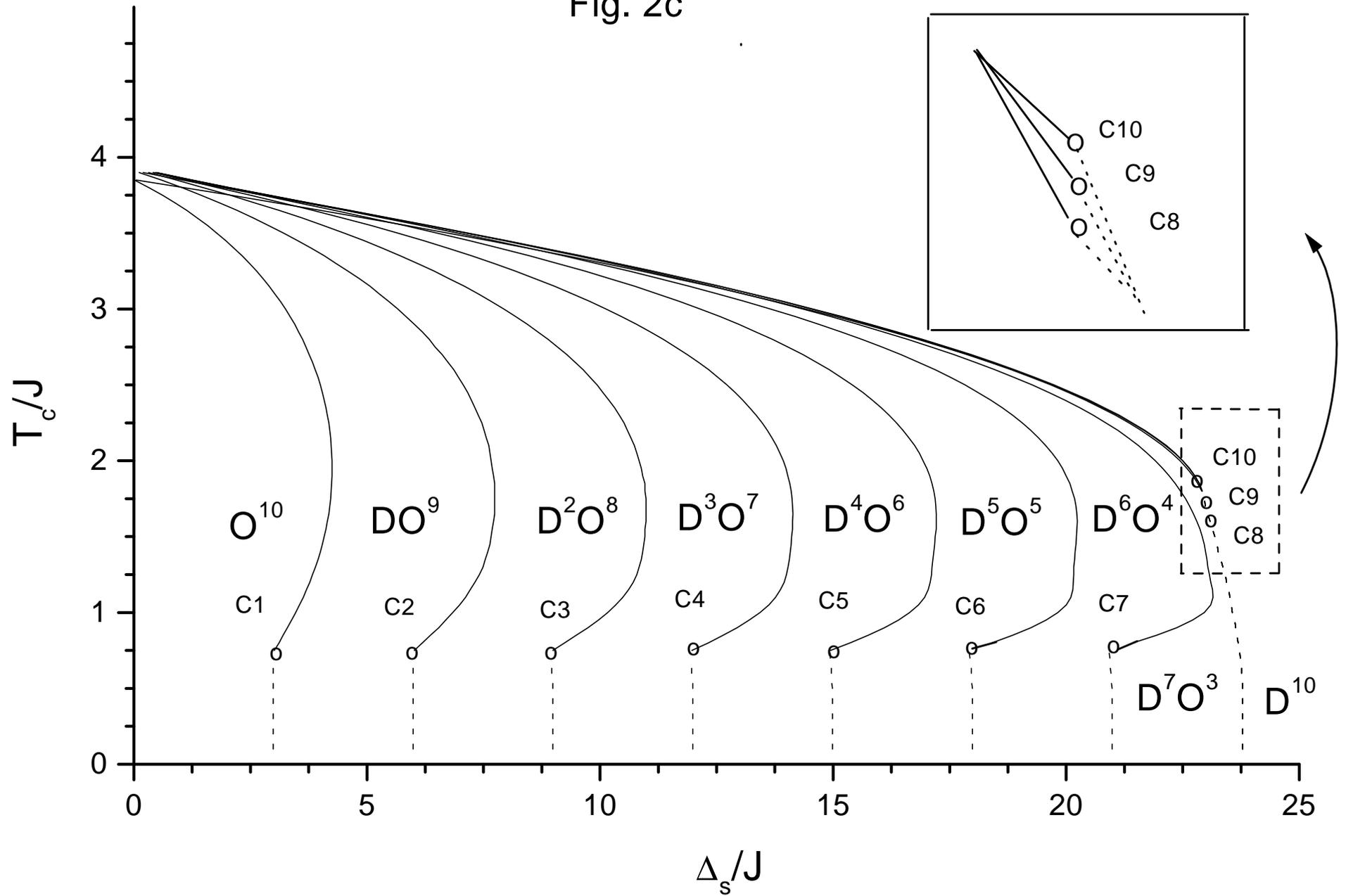
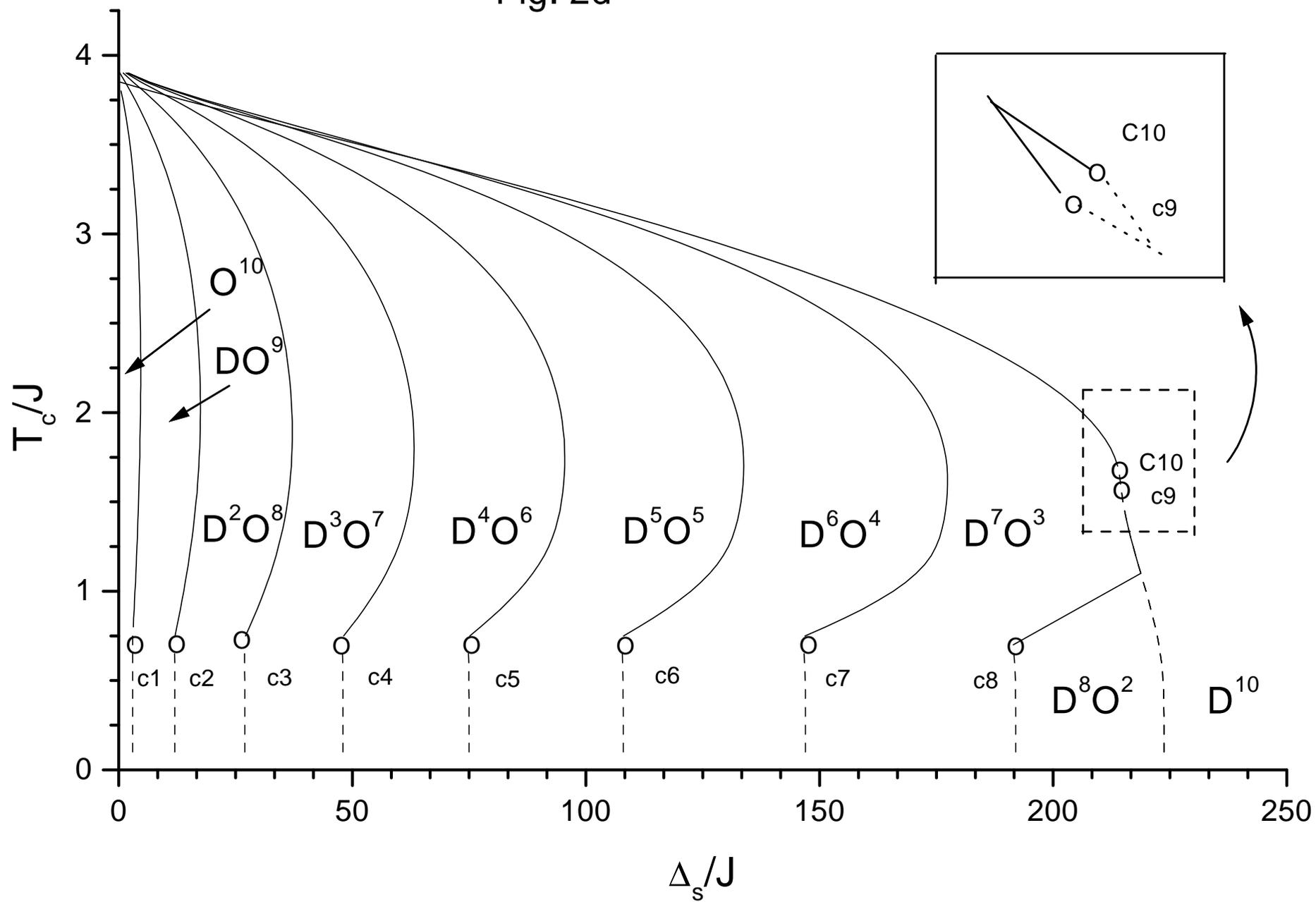
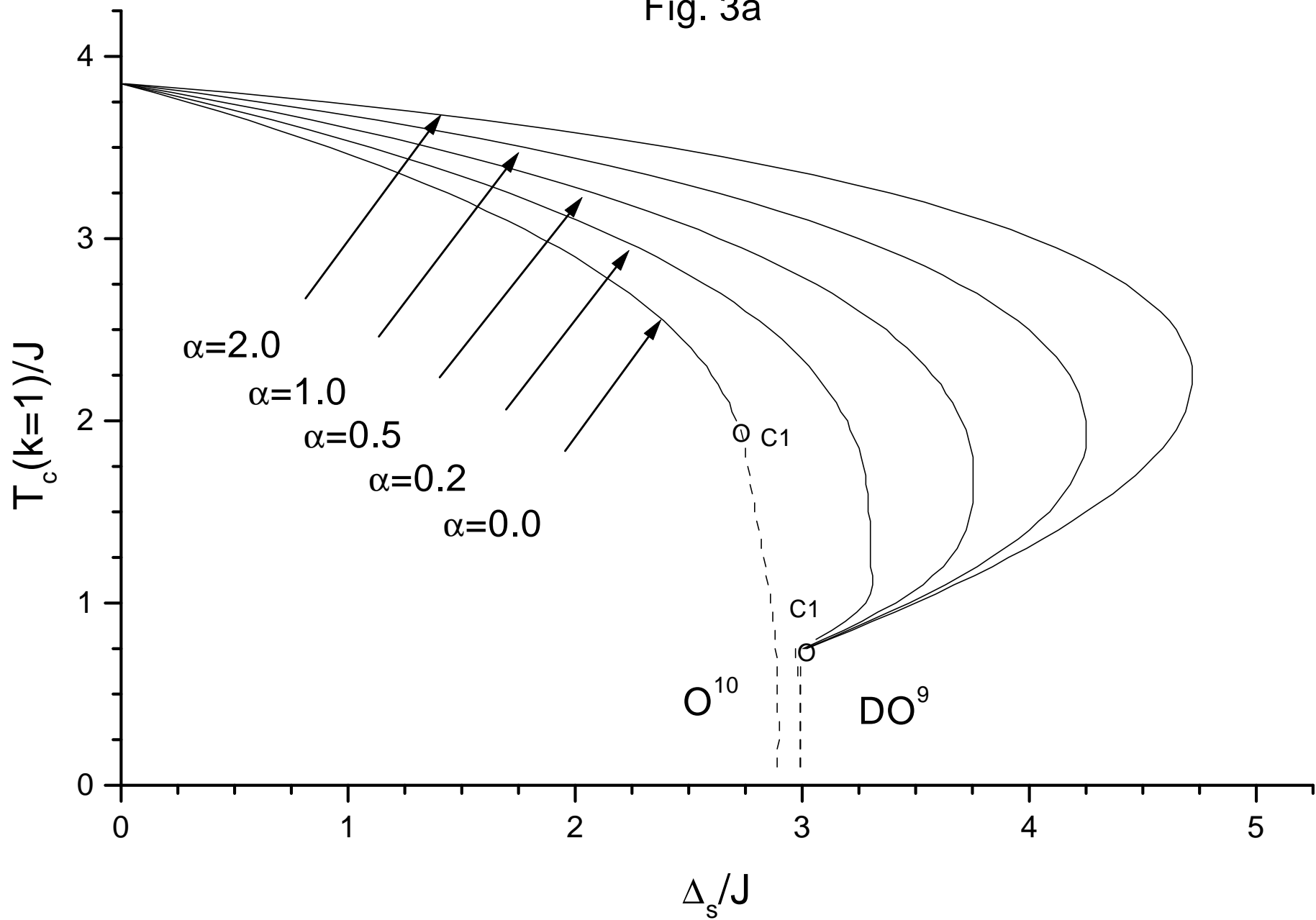
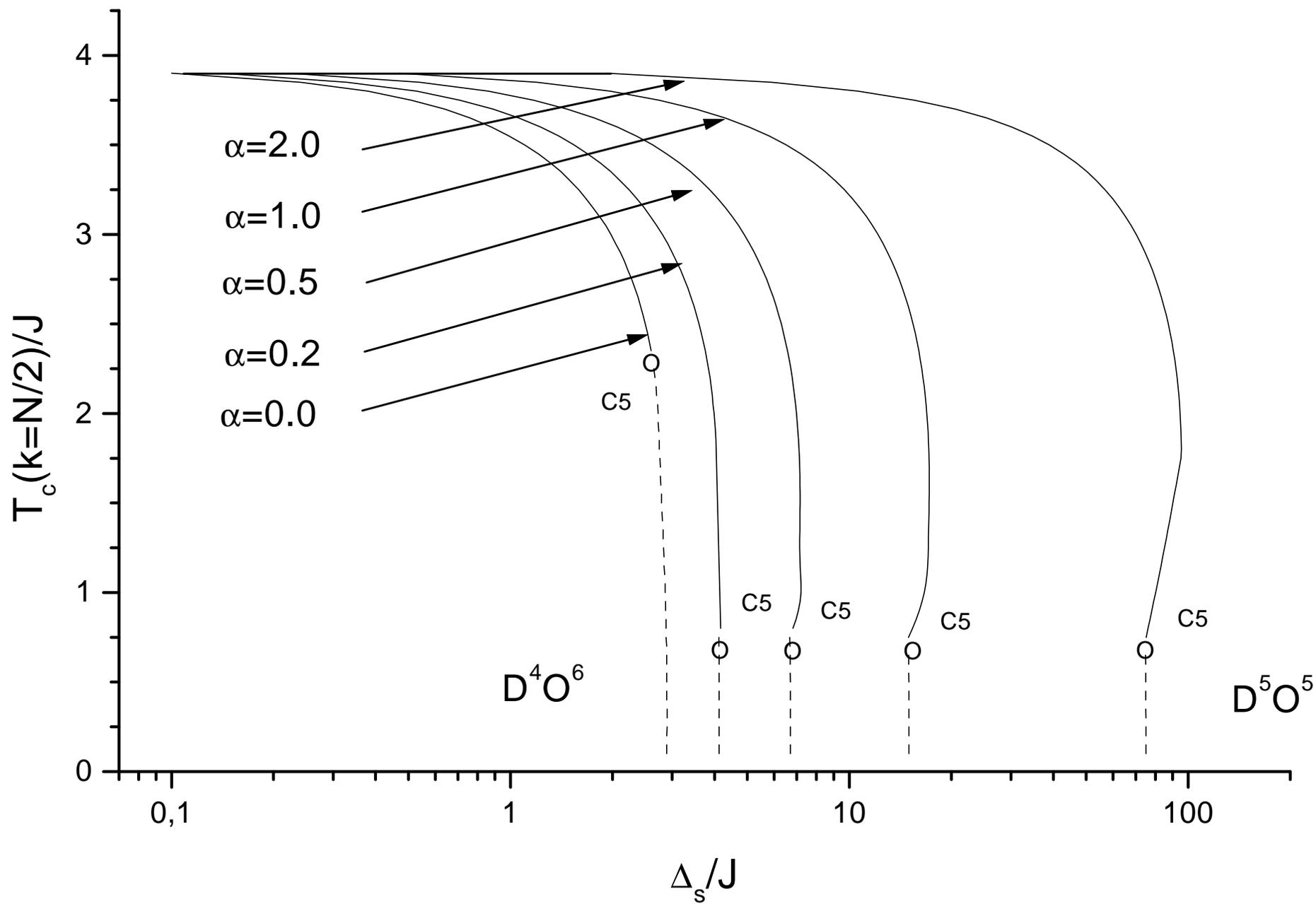
\end{document}